\documentclass[]{spie}  

\usepackage{amsmath,amsfonts,amssymb}
\usepackage{graphicx}
\usepackage[colorlinks=true, allcolors=blue]{hyperref}
\usepackage{booktabs}

\title{Region of interest detection for efficient aortic segmentation}

\author[a,b]{Loris Giordano}
\author[a,c]{Ine Dirks}
\author[b,d,e]{Tom Lenaerts}
\author[a,c,f]{Jef Vandemeulebroucke}

\affil[a]{Vrije Universiteit Brussel (VUB), Department of Electronics and Informatics (ETRO), Brussels, Belgium}
\affil[b]{Interuniversity Institute of Bioinformatics in Brussels, Brussels, Belgium}
\affil[c]{imec, Leuven, Belgium}
\affil[d]{Université Libre de Bruxelles (ULB), Machine Learning Group, Brussels, Belgium}
\affil[e]{Vrije Universiteit Brussel (VUB), Artificial Intelligence lab, Brussels, Belgium}
\affil[f]{Universitair Ziekenhuis Brussel (UZ Brussel), Department of Radiology, Brussels, Belgium}

\authorinfo{For further author information, send correspondence to Loris Giordano: loris.giordano@vub.be}

\begin{document}

\maketitle

\begin{abstract}

\label{sec:abstract}

Thoracic aortic dissection and aneurysms are the most lethal diseases of the aorta.
The major hindrance to treatment lies in the accurate analysis of the medical images.
More particularly, aortic segmentation of the 3D image is often tedious and difficult.
Deep-learning-based segmentation models are an ideal solution, but their inability to deliver usable outputs in difficult cases and their computational cost cause their clinical adoption to stay limited.
This study presents an innovative approach for efficient aortic segmentation using targeted region of interest (ROI) detection.
In contrast to classical detection models, we propose a simple and efficient detection model that can be widely applied to detect a single ROI.
Our detection model is trained as a multi-task model, using an encoder-decoder architecture for segmentation and a fully connected network attached to the bottleneck for detection.
We compare the performance of a one-step segmentation model applied to a complete image, nnU-Net and our cascade model composed of a detection and a segmentation step.
We achieve a mean Dice similarity coefficient of $0.944$ with over $0.9$ for all cases using a third of the computing power.
This simple solution achieves state-of-the-art performance while being compact and robust, making it an ideal solution for clinical applications.

\end{abstract}


\section{INTRODUCTION}

\label{sec:introduction}

\subsection{Clinical context}

The aorta is the main blood vessel of the human body, transporting blood from the heart to all the other organs.
This critical structure is subject to the greatest intravascular pressure changes throughout the cardiac cycle, making it vulnerable if its wall strength is affected by medical problems, genetic conditions, or trauma.
The most critical thoracic aortic conditions encompass thoracic aortic dissections and thoracic aortic aneurysms \cite{goldfinger2014thoracic}.
Both are characterized by local increases in aortic size, leading to distorted geometries and, potentially, aortic rupture, causing life-threatening internal bleeding. 
Measurement of aortic diameter on medical images to detect dilatation early-on is the primary approach to diagnose such conditions and support treatment decisions.

\subsection{Aortic segmentation models} 

Accurate measurement of the aorta in medical images is instrumental in the optimal treatment of thoracic aortic diseases.
Since this is still often done manually, obtaining high-quality results is time-consuming, labor-intensive, and subject to interpretation bias. 
Increasingly, clinicians opt for semi-automated and automated aortic analysis, which requires prior segmentation.

Previously, a variety of different approaches have been proposed for aortic segmentation, ranging from model-based solutions to wavelet analysis \cite{pepe2020detection}. 
Such models tend to perform well for healthy anatomies, but the accuracy for dissections and aneurysms plummets \cite{pepe2020detection}. 
Additionally, robustness and efficiency receive less attention but are critical to ensure that models are accepted and actively used in a clinical context.
More recently, deep-learning-based methods have received increased interest due to their performance and wide range of applicability for semantic segmentation.
Li et al.\cite{li2019lumen}, Cao et al.\cite{cao2019fully}, and Cheng et al.\cite{cheng2020deep} introduced deep segmentation models for dissected aortas, reaching Dice similarity coefficients (DSC) of around $0.91$. 
Wodzinski et al. \cite{wodzinski2023automatic} showed that with access to very large computational power, DSC of over $0.94$ could be obtained, claiming the first place in the Seg.A aorta segmentation challenge 2023 \cite{pepe2024segmentation}. 

Deep-learning often comes at the cost of high computational requirements, especially in the case of images with a large field of view, such as scans of the aorta.
Methods like 2D slice-based models or patch-based training and inference are usually adopted to mitigate the high computational requirements.
Processing an image in a piecewise manner reduces the amount of information that the model needs to process simultaneously but forces the model to make predictions using only a small portion of the large field of view.
This loss of context becomes more significant as the size of the pieces (slice or patch) becomes small compared to the field of view.

\subsection{Context in large images}

The most common approach to segmenting structures in images with a large field of view is to perform sliding window inference.
When the patch size is too small compared to the image size, the model receives too little context and produces results that are often underwhelming.
The state-of-the-art nnU-Net \cite{isensee2021nnu} proposes a cascade (low to high-resolution) network to mitigate this issue and provide context during the final segmentation.

Recently, region of interest (ROI) detection has been widely adopted to reduce the size of the image without losing context \cite{cui2023improved}.
Faster-RCNN \cite{ren2016faster} and nnDetection \cite{baumgartner2021nndetection} are highly flexible solutions to detect multiple objects of different classes. 
For our application of single organ detection, this flexibility is not needed and the models are way too complex.

\subsection{Contributions}

In this work, we explore ROI detection further to propose a compact method to train a single-organ segmentation model from images with a large field of view. 
We propose a compact detection model by condensing complex state-of-the-art models.
Additionally, we introduce a novel detection head to better comply with the requirements of our task.
The final output is obtained by cascading our detection model with a focused segmentation model.

We aim to explore the benefits of presenting an aortic segmentation model with more focused images on output quality and computational power.
We compare our approach to nnU-Net and a one-step segmentation model applied to the complete image regarding segmentation performance and computational resources.
The aim of our pipeline is to decrease the computing power needed for training and inference without impeding the final output quality.
We show that our approach achieves state-of-the-art aortic segmentation results while being simple, efficient, and transferable. 

\section{MATERIALS AND METHODS}

\label{sec:material_methods}

\begin{figure}[h]
\centering
\includegraphics[width=0.954\textwidth]{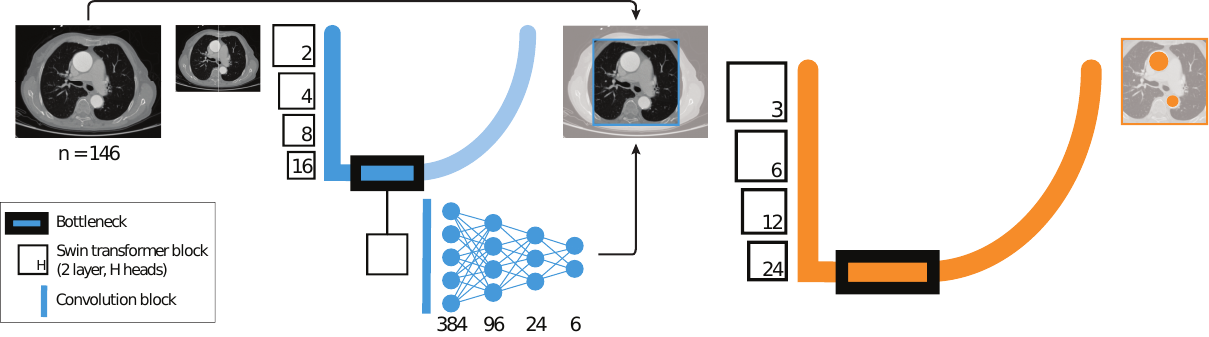}
\caption{Overview of our novel approach. We perform region of interest detection followed by focused segmentation of the cropped image to maximize output quality and minimize computational resources.} 
\label{fig:overview}
\end{figure}

\subsection{Data}

All data used in this study were retrieved from open-access databases.
Computed tomography (CT), often with contrast enhancement, is the predominant imaging modality for assessing aortic diseases.
We identified two datasets that can be used for this study.
First, the \textit{Aortic vessel tree (AVT)} \cite{radl2022avt} dataset was available for the MICCAI Seg.A challenge and contains $56$ patients with pixel-level annotations. 
Most images are labeled 'healthy', but several cases of dissections are present.
Second, the \textit{ImageTBAD} \cite{yao2021imagetbad} dataset is an open-source dataset containing $100$ patients with pixel-level annotations. 
All patients show some degree of type B aortic dissection (TBAD), i.e., a dissection in the descending part of the thoracic aorta, in contrast to type A aortic dissections (TAAD) found in the ascending part.
More details on both datasets are available in Table \ref{tab:datasets}.

\begin{table}[h!]
\centering
\caption{Overview of the data.}
\label{tab:datasets}
\begin{tabular}{lcc}
\toprule
 & \textbf{AVT} & \textbf{ImageTBAD}\\
\midrule
Number of patients & $56$ & $100$ \\
Imaging modality & CT & CT \\
Pathologies & Healthy ($50$) & Type B aortic dissection ($100$) \\
 & Aortic dissection ($5$) & \\
 & Aortic aneurysm ($1$) & \\
 \midrule
Image size (sagittal $\times$ coronal) & $512\times(512-666)$ & $512\times512$ \\
Image size (axial) & $94$ - $1140$ & $135$ – $416$ \\
Axial slice thickness & $0.5$ - $5$ mm & $0.4$ - $3$ mm \\
\bottomrule
\end{tabular}
\end{table}

Although both datasets are well defined and openly available, several steps are needed to blend them together.
First, we exclude all images with a resolution less than $2$ mm in the \textit{ImageTBAD} dataset.
This exclusion is not applied to the \textit{AVT} dataset, since applying the same strategy would remove $29$ healthy images from the $50$ available.
We deem that the added value of incorporating those images outweighs the problems of upsampling in the axial direction.
This leaves us with $146$ images from the initial $156$.
Then, all images are resampled to an isotropic $2$ mm resolution. 
The choice of $2$ mm is not ideal since the loss of detail compared to $1$ mm resolution is important, but only $62\%$ of our database has a sub-millimeter resolution.

The images are then padded or cropped to a fixed size of $256\times256\times256$ according to the following rules.
In the transverse dimensions (sagittal and coronal), images are center-cropped to $256$ voxels.
The images with transverse dimensions that are under $256$ voxels are border-padded symmetrically to $256$ voxels.
In the axial dimension, images see their inferior (i.e., away from the head) part cropped to a dimension of $256$ voxels. 
This results in the removal of the most inferior part of the abdominal aorta, which is not the focus of our research. 
The images with an axial dimension that is under $256$ voxels are then border-padded to $256$ voxels.
Finally, images using a Hounsfield units (\textit{HU}) definition of the intensity (\textit{I}) see their voxel values transformed into positive values according to $I = I_{HU} + 1024$.

\subsection{Computational setup}

We primarily use a Pascal architecture with $8$ Nvidia Tesla P100 with $16$ GB GPU memory. 
We employ an Ampere architecture with $10$ Nvidia Tesla A100 with $40$ GB GPU memory for the larger models.

\subsection{Detection model}

\subsubsection{Ground-truth labels}

Obtaining ground-truth labels is often one of the most difficult and time-consuming tasks when training a deep-learning model.
Additionally, this step requires the utmost rigor and expertise to create the labels repeatably and accurately.
To facilitate this step, we propose a fully automatic generation of ground-truth bounding boxes based on the available ground-truth pixel-level annotations.
The resulting labels are validated by human experts.

\begin{figure}[h]
\centering
\includegraphics[width=0.9\textwidth]{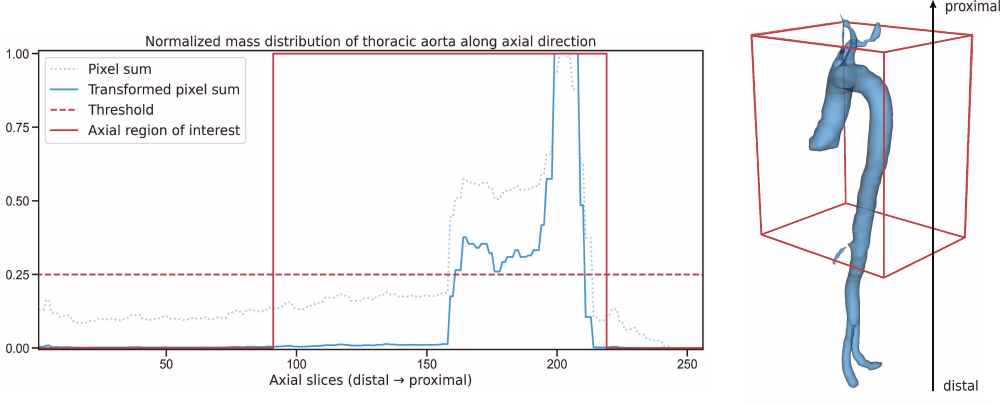}
\caption{Region of interest extraction by applying a threshold on the normalized aortic mass along the axial direction. A non-linear transformation is applied to exaggerate the difference between 'heavy' and 'light' slices.} \label{fig:pixel_sum}
\end{figure}

Our approach stems from the observation that the number of voxels representing the aorta is unequally distributed along the axial direction.
Figure~\ref{fig:pixel_sum} displays the normalized pixel sum (PS) found in axial slices against the axial slice number.
The most striking observation is that the PS plummets towards the most superior slices.
In the following reasoning, we substitute the PS with the term 'mass' for intuition, making the simplifying assumption that the mass is linearly correlated with the number of voxels.

Indeed, Figure~\ref{fig:pixel_sum} can be understood intuitively by several facts.
First, the diameter of the abdominal part of the aorta is smaller than the thoracic part in normal circumstances.
More importantly, the most superior part of the thoracic aorta has an ascending and descending part that run more or less parallel in the axial direction.
This approximately doubles the amount of mass in the superior part of the thoracic aorta.
Finally, the aortic arch has even more mass than the ascending and descending parts it connects.
Above the aortic arch, only the three aortic branches contribute to the mass, explaining the sudden drop towards the most superior slices.

Using these observations, the last slice containing the aortic arch can easily be identified.
We use a non-linear transformation to exaggerate the difference in mass between 'heavy' and 'light' slices, followed by a threshold operation.
From the identified slice, the start of the bounding box in the axial direction is given by the slice $20$ mm proximal to the identified slice. 
This extension in the bounding box is introduced to ensure that the roots of the aortic arch branches are in the ROI, since these are of great importance to expert radiologists.
The end of the bounding box in the axial direction is defined by considering the size of the bounding box in the axial direction, in this case 128 voxels.

For the sagittal and coronal dimensions, the center of the ROI is calculated as the center of mass in the respective directions. 
It is important to note that the sagittal and coronal center of mass are calculated in the axial ROI, discarding all other slices.
The ROI size in sagittal and coronal dimensions is given by the extents of the aorta in the respective directions with a margin of $20$ mm in each direction.

\subsubsection{Detection model}

Our observation is that available detection models like nnDetection and Faster-RCNN are designed for the challenging task of detecting several objects at different scales.
For our application of single-organ detection, we design a simple detection head to position one bounding box around the organ of interest. 
We base our architecture and training pipeline on the state-of-the-art nnDetection.

\paragraph{Complexity reduction}
nnDetection uses the Retina U-net architecture, composed of an encoder-decoder performing the segmentation task and doubling as a feature pyramid network to which several detection heads are attached.
Detection heads are composed of a classification head and a regression head.
The first reduction in complexity comes from the complete trimming of the classification heads required to identify the class of the bounding boxes.
Additionally, our regression head is built to position only one anchor with a given size and aspect ratio. 
For our application, we set the anchor size to $0.5\times0.5\times0.5$, using a cubic shape for simplicity.
Finally, following the reasoning of nnDetection, our simplified detection head only takes the feature map extracted from the bottleneck containing the highest-level semantic features.
The main reason for considering feature maps of other resolutions, i.e., identifying smaller objects, falls outside of the scope of this model.

\paragraph{Detection head}
We take advantage of our simplified task to design a small yet powerful detection head based on a fully connected network (FCN).
The output of the bottleneck still contains a large number of features ($16E \times \frac{p_1}{32} \times \frac{p_2}{32} \times \frac{p_3}{32}$ with \textit{E} the size of the embedding space and \textbf{\textit{p}} the patch size).
We first condense the output of the bottleneck by applying a convolution block with a kernel of identical size to the spatial size of the bottleneck and twice the number of channels.
This operation allows us to obtain an even more compact representation of the scene while flattening the feature map using the spatial relations between the features.
The convolution is followed by layer normalization, parametric rectified linear unit (ReLU) activation, and a squeezing operation to obtain a flat list of features ($2 \cdot 16 \cdot E$ with E the size of the embedding space).
The very compact representation is then passed through a FCN with an input size equal to the number of features and an output size of $A \times 2 \times 3$ with A the number of anchors to be detected in the image.
For the case of single-organ detection, A is equal to $1$.
For each anchor, the first $3$ outputs represent the position in the image, while the last $3$ outputs represent the scaling of the predefined anchor.
Two hidden layers are added between the input and output layers, with sizes given by the geometric sequence defined by the input and output size.
Each layer of the FCN is followed by sigmoid activation, ensuring a final output between $0$ and $1$. 
This approach essentially presents a given bounding box to the model with a certain aspect ratio and maximal size that can be positioned and scaled down to fit the organ of interest.

\begin{table}[h]
\centering
\caption{Hyperparameters for the detection model.}
\label{tab:detection_hyperparameters}
\begin{tabular}{lc}
\toprule
\textbf{Hyperparameter} & \textbf{Value/method} \\
\midrule
Input channels & $1$ (thoracic CT scan)\\
Output channels & $1$ (segmentation map)\\
Size of embedding space & $12$\\
Number of layers per stage & $2$\\
Number of attention heads & $2$ - $4$ - $8$ - $12$\\
Normalization & Instance\\
\midrule
Number of hidden layers & $2$\\
Size of hidden layers & Geometric sequence defined by $\frac{input\ size}{output\ size}$\\ 
Activation & Sigmoid\\
\midrule
Training procedure & $1500$ epochs, validation every $5$ epochs \\
Batch size & $2$ \\
Loss function & Generalized Interesection over Union (GIoU)\\
 & Combined Dice and Cross-entropy (DiceCE) \\
Learning rate & $0.01$ with Poly learning rate scheduler \\
Optimizer & SGD with Nesterov momentum ($0.99$) \\
Inference & Simple \\
Validation & Intersection over Union (IoU) \\
\bottomrule
\end{tabular}
\end{table}

\paragraph{Training and inference}

Similarly to nnDetection \cite{baumgartner2021nndetection}, we use a multi-task setup to train our detection model.
First, the images are downsampled until they can be processed as one patch using a batch size of $2$.
Considering our computational setup, we use images with an isotropic resolution of $4$ mm, corresponding to an image size of $128\times128\times128$.
We opt for a SwinUNETR \cite{hatamizadeh2021swin} instead of Retina U-net as backbone, with hyperparameters listed in Table \ref{tab:detection_hyperparameters}. 
In line with the overall reduction in complexity in our model, we chose a lightweight SwinUNETR architecture with a reduced number of attention heads and a small embedding space.
Even though originally designed for a U-Net architecture, we follow relevant guidelines of nnU-Net \cite{isensee2021nnu} to determine the training hyperparameters.
Most notably, mirroring is omitted for our application as some valuable anatomical information could be lost in this process for structures such as the aorta, which predominantly have a certain position and direction in the human body.
We adopt a very simple post-processing step for both segmentation and detection.
The segmentation output is obtained by passing the decoder output through a sigmoid activation function followed by a threshold at $0.5$.
The detection output is obtained by multiplying the detection head output with the image size.
Like nnDetection, the total loss is given by the sum of the Generalized Intersection over Union (GIoU) loss for detection and the combined Dice and Cross-entropy (DiceCE) loss for segmentation.
Figure \ref{fig:overview} shows an overview of our detection model.
An encoder-decoder structure performs the segmentation; a regression head is attached to the bottleneck of the segmentation model for detection.

\subsection{Segmentation model}

\subsubsection{Baseline: nnU-Net and one-step segmentation}

As a baseline model, we chose the state-of-the-art nnU-Net \cite{isensee2021nnu}.
We rigorously follow the nnU-Net pipeline using the maximal amount of computational power at hand.

Similarly to our cascade approach, we opt for a SwinUNETR \cite{hatamizadeh2021swin} for the segmentation model with hyperparameters listed in Table~\ref{tab:segmentation_hyperparameters}.
The number of parameters of this final segmentation model is increased compared to the detection model backbone (Table \ref{tab:detection_hyperparameters}) by increasing the number of attention heads and the size of the embedding space.
To compensate the increase in the number of parameters, the patch size is reduced accordingly.
This segmentation model serves as the baseline segmentation model operating on the complete image and the region of interest identified by the detection model (Figure~\ref{fig:overview}).

\begin{table}[h]
\centering
\caption{Hyperparameters for the segmentation model.}
\label{tab:segmentation_hyperparameters}
\begin{tabular}{lc}
\toprule
\textbf{Hyperparameter} & \textbf{Value/method} \\
\midrule
Input channels & $1$ (thoracic CT scan)\\
Output channels & $1$ (segmentation map)\\
Size of embedding space & $24$\\
Number of layers per stage & $2$\\
Number of attention heads & $3$ - $6$ - $12$ - $24$\\
Normalization & Instance\\
\midrule
Training procedure & $1500$ epochs, validation every $5$ epochs \\
Batch size & $4$ \\
Patch selection & $1/2$ ensured centered around foreground voxel\\
Patch size & $96\times96\times96$ \\
Loss function & Combined Dice and Cross-entropy (DiceCE) loss \\
Learning rate & $0.01$ with Poly learning rate scheduler \\
Optimizer & SGD with Nesterov momentum ($0.99$) \\
Inference & Sliding window inference with half overlap \\
Validation & Dice similarity coefficient (DSC) \\
\bottomrule
\end{tabular}
\end{table}

\subsubsection{Cascade detection \& segmentation}

To train the segmentation models operating on the regions of interest instead of the whole image, the same architecture and hyperparameters as the one-step segmentation pipeline are used.
In practice, the size of all the bounding boxes is extended to the maximal size ($128\times128\times128$) before the cropping operation is applied, adding an extra layer of quality assurance while maintaining the homogeneity of the image size in our dataset. 
The detection and segmentation models are trained sequentially. 
The segmentation model is trained using cropped images obtained via immediate inference of the detection model.

\subsection{Validation \& Metrics}

We opt for a train-validation-test split ($70\%$-$15\%$-$15\%$), stratified by type of pathology, with at least one representation of each pathology in the test and validation set. 
For the only abdominal aortic aneurysm present in the \textit{AVT} dataset, it is treated as if it was a healthy patient since this pathology is represented only once, and the abdominal aorta falls outside of the scope of our research. 
The chosen validation metrics are selected based on the relevant literature.

\begin{itemize}
    \item \textit{Dice similarity coefficient (DSC)} is the main reported metric in other work related to the segmentation of dissected thoracic aortas. 
    The DSC of state-of-the-art approaches hover around $0.91$ \cite{cao2019fully, li2019lumen, cheng2020deep}. 
    Recently, Wodzinksi et al. obtained a DSC of over $0.94$ \cite{wodzinski2023automatic}. 
    Considering inter-observer variability (for human experts), this is as close as one can get to perfect segmentation results \cite{montagne2021challenge}.
    In general, output segmentation with a DSC over $0.8$ could be considered for clinical validation\cite{morozov2019clinical}. 
    All our final output segmentation maps are evaluated using DSC.
    \item \textit{Intersection over union (IoU)} is used as validation metric for the detection models. 
    Good IoU values are considered to be higher than $0.7$ \cite{baumgartner2021nndetection, zhang2021loss}. 
    Our detected bounding boxes are evaluated using IoU. 
    \item \textit{Complete containment (CC)} is a criterion not reported in the literature that we introduce here as an additional check to ensure the practical quality of the detection. 
    Indeed, to be useful in practice, the ROI must always contain the organ of interest completely, in this case, the superior thoracic aorta. 
    To perform this check, the ground-truth bounding boxes are trimmed to remove the margins, and it is ensured that the predicted bounding box contains at least the trimmed ground-truth.
\end{itemize}

\section{RESULTS}

\label{sec:results}

\subsection{Detection}

\begin{figure}[h]
\centering
\includegraphics[width=\textwidth]{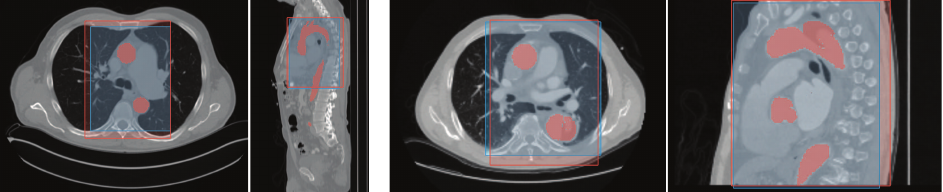}
\caption{Two examples of the detection results. Red indicates the ground truth bounding box and segmentation. Blue indicates the predicted bounding box. The mean IoU is over 0.8, with complete containment of the thoracic aorta for each image.} \label{fig:detection}
\end{figure}

Figure~\ref{fig:detection} shows a visual comparison of the output of our detection model together with the ground truth.
The results of our detection model are summarized in Table \ref{tab:detection_results}.

\begin{table}[h]
\centering
\caption{Results for aortic detection on the test set.}
\label{tab:detection_results}
\begin{tabular}{lc}
\toprule
\textbf{Metric} & \textbf{Value} \\
\midrule
IoU (mean $\pm$ std) & $0.806 \pm 0.075$ \\
DSC  (mean) & $0.918$ \\
Complete containment & $100\%\ (146/146)$\\
\midrule
Training GPU memory [GB] & $10.7$ \\
Inference GPU memory [GB]  (max) & $2.5$ \\
Training time [h] & $19.3$ \\
Inference time [s] (mean) & $0.146$ \\
\bottomrule
\end{tabular}
\end{table}

The mean IoU of this approach for the validation set is $0.806$, corresponding to a mean difference of $2.38$ slices in all dimensions compared with the ground truth.
The mean DSC for the segmentation part of this multi-task model is $0.918$. 
Regarding complete containment, this method produces bounding boxes that always contain the ROI for our complete dataset.
The required computational resources to train this model with the presented data and hyperparameters is a GPU with at least $10.7$ GB for a duration of $19$ hours.
The mean inference time is $0.146$ seconds (not taking into account any downsampling), and the maximal required GPU memory is $2.5$ GB.

\subsection{Segmentation}

\begin{figure}[h]
\centering
\includegraphics[width=\textwidth]{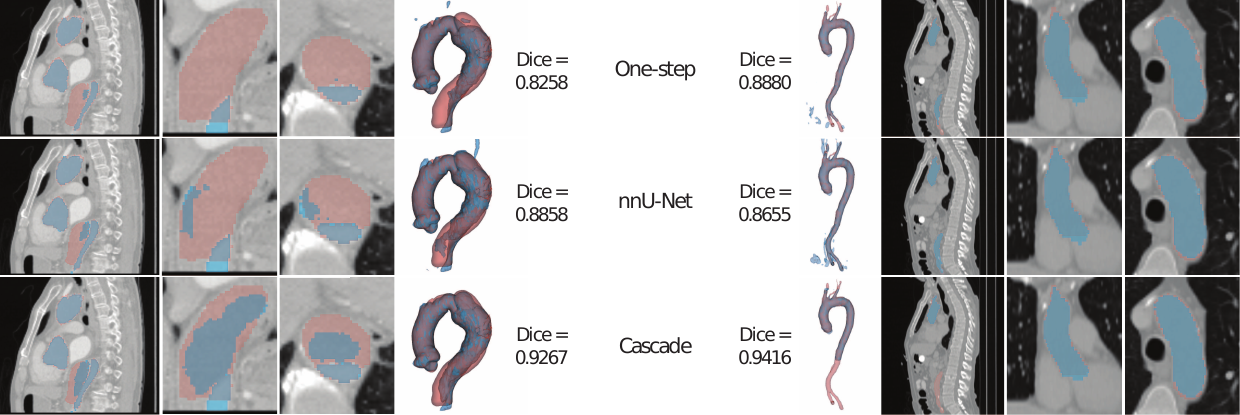}
\caption{Two examples of the segmentation results using the one-step, nnU-Net, and cascade approaches. The results are shown in the three main anatomical directions (from left to right: sagittal, coronal, axial) and as a 3D view of the segmentation. Red indicates the ground truth. Blue indicates the output segmentation. The mean Dice is the highest for the cascade approach, with values well over 0.9 and clear visual improvements.} \label{fig:segmentation}
\end{figure}

Figure~\ref{fig:segmentation} shows some results of the different segmentation models.
Table~\ref{tab:segmentation_results} shows the performance of our cascade model compared to our one-step segmentation model and nnU-Net.

\begin{table}[h]
\centering
\caption{Results for aortic segmentation on the test set.}
\label{tab:segmentation_results}
\begin{tabular}{lccccc}
\toprule
\textbf{Metric} & \textbf{One-step} & & \textbf{nnU-Net} & & \textbf{Cascade} \\
\midrule
DSC (mean $\pm$ std) & $0.895\pm0.043$ & & $0.905\pm0.025$ & & $\mathbf{0.944\pm0.028\ }$ \\
HD95 [mm]  (mean $\pm$ std) & $5.685\pm5.500$ & & $4.342\pm3.125$ & & $\mathbf{2.616\pm1.278}$ \\
Patch coverage & $5.1\%$ & & $12.5\%$ & & $42.2\%$ \\
\midrule
Training GPU memory [GB] & $32.7$ & & $36.4$ & & $\mathbf{10.3}$ \\
Inference GPU memory [GB] (max) & $\mathbf{1.59}$ & & $2.03$ & & $2.55$ \\
Training time [h] & $25.3$ & & $27.0$ & & $\mathbf{18.5}$ \\
Inference time [s] (mean) & $6.20$ & & $0.76$ & & $\mathbf{0.61}$ \\
\bottomrule
\end{tabular}
\end{table}

Wilcoxon signed-rank tests on DSC and HD95 reveal that the performance of our proposed method is significantly better than the one-step method (DSC:~$p < 0.001$, HD95:~$p < 0.001$) and nnU-Net (DSC:~$p < 0.001$, HD95:~$p < 0.001$).
It is important to note that all the outputs of our cascade model have a DSC over $0.9$.

Interesting to mention is the improved performance of nnU-Net and our cascade model in terms of the standard deviation of the DSC.
Indeed, the one-step approach shows almost double the standard deviation on DSC compared to the two other methods.
When looking at individual results, the one-step approach occasionally outputs segmentations with low DSCs of around $0.8$, which is not the case for the other methods.

The GPU memory required to train the cascade model is the lowest, using only $10.3$ GB, while the patch coverage is the highest, with more than $40\%$ of the image being processed at once.
It is important to note that the training time for the cascade model does not contain the training time for the detection model.  

\section{DISCUSSION}

\label{sec:discussion}

\subsection{Detection}

The detection model performs well (mean IoU $> 0.7$) with no measured qualitative influence on the capability of the model to identify the ROI (complete containment).
It is important to note the simplicity of the detection head, composed of one learnable convolution block followed by a small fully connected network.
Moreover, having access to high-level features is the only requirement of our detection head, making it compatible with other encoder-decoder architectures.

\subsection{Segmentation}

Our cascade model significantly outperforms our one-step segmentation model and nnU-Net trained on the same dataset regarding segmentation quality metrics.
All output segmentations of our cascade model have a DSC of over $0.9$.
Moreover, the standard deviation of the DSC and HD95 is twice as low as for the other methods, and outputs for complex cases are always of high quality, showing the increased robustness of our method for complex cases.
We attribute the increased robustness to the larger patch coverage ($> 12.5\%$ \cite{isensee2021nnu}) and the more focused images provided by our approach.
This allows our final segmentation model to learn to segment only relevant parts of the image in detail.
Additionally, our cascade approach requires less than a third of the memory for training, considerably lower training time, and has a mean inference time of less than one second.
The required GPU memory for inference is the highest of the three methods but still low enough that it could be run on a modern computer.
High output quality, robustness, and low computational requirements are instrumental for a model that can be used actively, maintained, and fine-tuned further in a clinical setting.

Figure~\ref{fig:segmentation} shows the two mechanisms that help the final segmentation model of our cascade approach produce more accurate results more reliably: focus and image size.
On the left, the raw image has a small field of view around the aorta.
In this case, the region of interest is large compared to the image, and the detection step mostly centers the image around its most interesting part without dramatically reducing its size.
The increased performance of our cascade approach compared to the one-step approach and nnU-Net is attributed to the more focused training based on ROIs instead of the whole images.
This stems from the fact that the final segmentation of our cascade model only learned relevant features inside the ROI without needing to dedicate parts of the model to less interesting features outside of the ROI.
On the right, the raw image is large compared to the ROI, leading to the detection step not only centering the image around its most interesting part but also reducing the overall image size.
In addition to the focused view of the aorta, the reduction in size also allows our final segmentation model to be applied with a larger patch coverage using the same computational power.
This leads to the details being handled slightly better by our cascade model than the other approaches.
Additionally, the detection step also allows the cascade model to completely neglect the lower part of the image, where additional errors can occur.
This is especially true in this case, where the iliac arteries are handled poorly by both the one-step model and nnU-Net, likely due to a lack of representation of these types of images in our dataset.

Our approach outperforms state-of-the-art approaches focused on dissected aortas \cite{cao2019fully, li2019lumen, cheng2020deep} (mean DSC $>0.91$) and compares to the best-performing model of Seg.A aorta segmentation challenge \cite{pepe2024segmentation, wodzinski2023automatic} (mean DSC $= 0.943$, mean HD95 $= 1.90$ mm).
It is important to note that our dataset is represented by a majority of cases reported to be more challenging\cite{pepe2020detection} (100 dissections in \textit{ImageTBAD}).
On the contrary, Wodzinski et al.\cite{wodzinski2023automatic} only had a third of the data to train their model.
While our model is trained, validated, and tested on more general and more challenging data, it cannot be excluded that the large 3D model presented by Wodzinski et al. could benefit from a large increase in training samples.

We cannot directly compare training resources of other methods.
Given the setup reported by Wodzinski et al. \cite{wodzinski2023automatic} (PLGrid infrastructure), we assume the required GPU memory for their approach to be 96 GB.
However, inference memory and time are usually reported.
Wodzinski et al. report an inference time of less than 2 seconds using an identical computational setup.
Cao et al. and Cheng et al. report inference times of $6.28$\cite{cao2019fully} and $5.35$\cite{cheng2020deep} seconds, respectively, using comparable but less powerful GPUs.

\subsection{Clinical advantages}

The computing resources required to train our detection and segmentation model are limited enough so that the results can be reproduced without requiring access to a supercomputer cluster. 
This could open the door to fine-tuning the model after deployment. 
By allowing clinicians to make (small) adaptations to the output bounding boxes and the output segmentation maps, new ground-truth annotations could be fed back to our model and used for training on a local GPU.

The tools presented in this research could also be used as a basis for a quality assurance pipeline.
Our automatic generation of the ground-truth labels could be easily applied to the low-resolution output segmentation of the detection model to serve as an additional check on the position of the bounding box.
Additionally, the same low-resolution segmentation could be used as a preview of the aortic segmentation of parts that fall outside the ROI.

Finally, using lightweight models to obtain state-of-the-art results is also an enormous advantage considering real-life societal problems. 
Medical doctors have ever-decreasing time for each patient that they need to share between analyzing the patient's data and interacting with the patients to understand their needs.
Having a rapid way to segment medical images can be a great support for medical doctors, but it is also a direct gateway to automated measurement taking, shape analysis, and predictive modeling.
Recent deep learning models are able to process and generate ever-increasing quantities of data at the cost of financial and environmental investments.
Having a small model that is able to produce focused and compressed outputs can reduce the need for such resources.

\subsection{Limitations}

The main limitation of this study is the relatively low resolution of the final segmentation model. 
Given the available datasets, we chose to work with lower-resolution images instead of oversampled images.
We are currently collecting more data with better resolution and foresee that this will improve our models even more.
The limited computational resources needed to achieve the current low-resolution model ensure that a high-resolution model can be trained using the same pipeline.

Another limitation is the absence of fine-tuning of the architecture and post-processing.
Standard hyperparameters have been used for the SwinUNETR architectures and the guidelines of nnU-Net have been followed when choosing training hyperparameters.
It is yet unknown if nnU-Net guidelines transfer well to SwinUNETR-based pipelines.
Additionally, elastic augmentations are not advised by nnU-Net but could be beneficial when considering a soft organ with high levels of deformations like a dilated or dissected aorta. 

Finally, the focus on the thoracic aorta could be a limitation for applications in which the complete aorta needs to be segmented at once.
One solution would be to redefine the bounding boxes in such a way that the complete aorta is considered inside the ROI.
This would still reduce the overall size of the image and allow the final segmentation model to see more focused images.
Another approach would be to consider a detection model for the thoracic and abdominal aorta cascaded by two independent segmentation models specialized in thoracic and abdominal segmentation, respectively.
This could still be a benefit in terms of output quality and computational power compared to the one-step approach.

\section*{DECLARATIONS}

\label{sec:declarations}

The work has not been and is not being submitted for publication or presentation elsewhere.

\acknowledgments 

\label{sec:acknowledgments}

This work has been funded by FARI - AI for the Common Good Institute in Brussels.

\bibliography{report} 
\bibliographystyle{spiebib} 

\end{document}